# Preliminary Results on a New Algorithm for Blink Correction Adaptive to Inter- and Intra-Subject Variability

E. Guttmann-Flury, X. Sheng, D. Zhang, and X. Zhu, *Member, IEEE*

*Abstract*— This paper presents a new preprocessing method to correct blinking artifacts in Electroencephalography (EEG) based Brain-Computer Interfaces (BCIs). This Algorithm for Blink Correction (ABC) directly corrects the signal in the time domain without the need for additional Electrooculogram (EOG) electrodes. The main idea is to automatically adapt to the blink's inter- and intra-subject variability by considering the blink's amplitude as a parameter. A simple Minimum Distance to Riemannian Mean (MDRM) is applied as the classification algorithm. Preliminary results on three subjects show a mean classification accuracy increase of 13.7% using ABC.

## I. Introduction

Brain Computer Interfaces (BCI) are gaining more and more attention from researchers and companies. Applications ranging from virtual reality to brain-typing are being developed and aspire for accessibility to a global public. Such systems need real-time interaction through brain activity without invasive implants.

BCI require the acquisition of small potential signals, in essence highly sensitive to artifacts. Particularly, non-invasive electroencephalography (EEG) is a challenging problem, since recorded signals are very low compared to biological artifacts. Specifically, oculomotor activity heavily contaminates the data by causing a change in the electrical fields that surround the eyes [1]. Furthermore, the instruction to refrain from blinking may seriously distort brain activity [2].

Many efforts have been made to understand and correct these effects [3,4]. Typically, a human eye blink has an amplitude which can be 10 to 100 times larger than the electrical signals originating from the cerebral cortex and can last up to 200ms to 400ms [5]. This artifact is primarily picked up by frontal electrodes but extends further [6]. The frequent occurrence of blinks represents a major source of artifacts in the EEG. Hence, several methods have been developed to cope with this nuisance.

Regression based methods estimate the relation between one or several reference Electrooculogram (EOG) channels and each EEG channel. With the premise that the reference channels properly represent all interference waveforms, artifacts can be corrected by subtracting a regressed portion of each reference channel from the contaminated EEG. This method main hypothesis is that: "*If the blinks do not vary in their amplitude and duration, the EOG represents a good estimate of the "true" blink signal*" [3]. Therefore, the regression error will be small. More recently, Independent Component Analysis (ICA) have been reported to be successful at removing artifacts from the EEG and has mostly replaced other approaches. This method allows to separate components in complex signals with the possibility of discriminating between artifacts and brain waves. The former is removed from the components and the signal is then reconstructed without it. Due to ICA being based on statistical features, results will not be reliable if the amount of data given to the algorithm is insufficient. Another drawback is the need of an "expert" to identify which component represents the ocular artifacts. Finally, the computational load is very high [3].

We conducted a first analysis on the online available BCI Challenge dataset from the IEEE Neural Engineering Conference (NER2015) [7]. Blinks from 26 subjects were extracted and clustered into bins using the maximum amplitude on a frontopolar channel (FP1) as a parameter. The blinks distribution analysis showed a high variability across subjects, thus highlighting the necessity of an adaptive physiology-based algorithm. Furthermore, our goal is to develop a "plug-and-play" blink artifact correction algorithm with low computational load and using only EEG channels. In this preliminary work, our Algorithm for Blink Correction (ABC) yielded promising results offline.

## II. Methodology

### A. Experimental Setup

Three healthy subjects participated in this experiment (2 males/ 1 female, all right-handed, average age of 32.3 years). This study was approved by the Ethics Committee of Shanghai Jiao Tong University, China. All participants signed informed consent form prior to participation. Furthermore, they were notified about the possibility of publishing any identifying picture.

During this paradigm, subjects are seated in a noise proof chamber, designed specifically for EEG recordings. A display is placed on a desk in front of the subject to provide visual instructions generated by the E-Prime software. Subjects were informed of the experimental protocol in advance.

EEG signals are recorded using a SynAmps2 system connected to an amplifier (Compumedics, Neuroscan). A 64 channel Quik-cap collects data from 62 EEG channels. The EEG electrodes are placed according to the extended 10/20 system with channel CZ situated on the vertex. The reference is located on the right mastoid and the ground electrode on the forehead. A high-pass filter at 0.1 Hz, low pass filter at 200 Hz, and a notch filter to remove the interference of the power supply at 50 Hz are applied to the data. The analog signal of all electrodes is sampled at 1000 Hz.

* This work was supported by in part the National Science Foundation of China (No.51620105002), and the Science and Technology Commission of Shanghai Municipality (No.17JC1402700).
E. Guttmann-Flury, X. Sheng, D. Zhang and X. Zhu are with the State Key Laboratory of Mechanical System and Vibration, School of Mechanical Engineering, Shanghai Jiao Tong University, 800 Dongchuan Road, Minhang District, Shanghai, 200240, P. R. China. (Corresponding e-mail: mexyzhu@sjtu.edu.cn).

During the experiment, the E-Prime software sends triggers to both the EEG amplifiers and a high-speed camera (Phantom M310) every 6 milliseconds. Such a setup allows for a comparison of the video to the corresponding EEG signal once a blink is detected.

One session is composed of five different tasks. In this preliminary work, the analysis is limited to the Motor Imagination (MI) of hand grasping (Left or Right). At the beginning of each trial, a white fixation cross appears in the center of the screen for 2 seconds. A red rectangle cue then appears randomly on the left or the right side of the cross for 4 seconds. Meanwhile, subjects perform kinesthetic motor imagination of grasping for 3 times. Finally, the fixation cross and the red rectangle cue disappear during a random rest of 1 to 1.5 second allowing relaxation while avoiding subject's adaptation. The experimental paradigm is illustrated in Fig. 1.

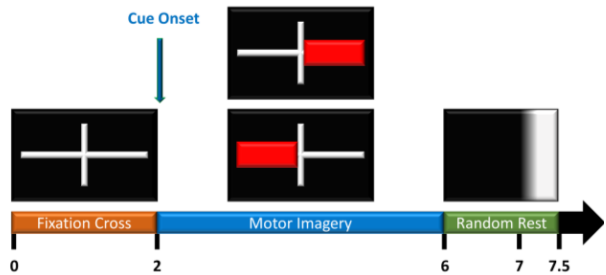

Figure 1. Experimental paradigm

This MI task lasts around 6 minutes per session with a total of 40 trials (20 Left + 20 Right randomly distributed). The task's order is also randomly chosen at the beginning of the session. A total of 3 sessions (120 trials) per subject is recorded on three different days.

At the beginning of each session, subjects had to fill in a questionnaire detailing their conditions at the time of recording. For example, the degree of alertness has been divided into four levels: "Rested", "Slightly tired", "Moderate fatigue" and "Extreme fatigue". At the end of each task, similar questions were asked.

### B. Algorithm for Blink Correction (ABC)

After downsampling and lowpass filtering the data with a 4[th]-order Butterworth filter, an amplitude threshold is applied on 2 seconds-wide windows over a reference frontopolar channel (FP1). The amplitude criterion selects all data higher than the moving average to which a heuristic threshold is added. For instance, if the reference moving average on the 2 seconds time window is 740 µV, any data points higher than 740 + 25 = 765 µV will be pre-selected. This heuristic threshold has been calculated from the comparative analysis between the EEG signal and the video of the eye. Most of the blink-free windows are already discarded after this first criterion.

On the remaining data, the first maximum amplitude is detected. The pre and post-amplitude criterion checks the blink shape likeliness, de facto removing saccades or electrode "pops". Followingly, propagation of the blink effect on a further electrode (CZ) confirms whether the selected data is a blink or a brain waveform.

If the selected data checked all three criteria, the blink is extracted. It is then classified into 20 µV-range bins, using its maximum amplitude as a parameter. Finally, Grand Averages (GA) are computed per subject and per bin.

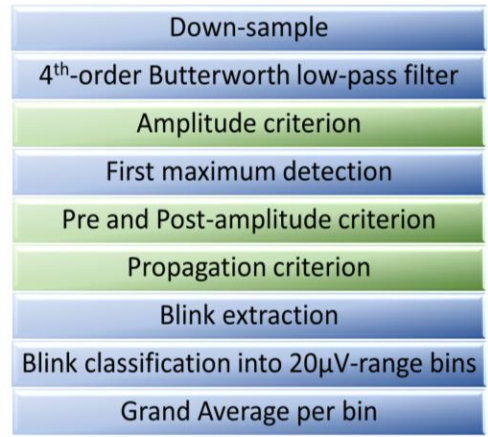

Figure 2. Algorithm for Blink Correction (ABC) flowchart

A succinct flowchart highlighting the main steps of ABC is presented in Fig. 2. In a nutshell, the correction process consists of three steps: (1) blink detection, (2) blink classification and (3) subtraction with the corresponding GA.

### C. Data Processing with Riemannian Geometry

The topographical representation and band power change of MI mental task are well-known. Different body parts are represented in different area of the motor cortex (right hand under C3 electrode, left hand under C4). Single-trial classification uses spatial covariance matrices to recognize the different mental tasks. Since the covariance matrices contain the spatial information embedded in EEG signal, most existing algorithms are based on their estimation.

Let $x_t \in \mathbb{R}^n$ denote the EEG signal vector at a specific time point $t$, with $n$ denoting the number of recording channels. Short-time segments of EEG signal, or trials, can be represented as a matrix $X_i = [x_{t+T_i} \ldots x_{t+T_i+T_s-1}] \in \mathbb{R}^{n \times T_s}$ which corresponds to the $i$-th trial of MI started at time $t = T_i$. For this trial, the spatial covariance matrix is estimated using the Sample Covariance Matrix (SCM) $P_i \in \mathbb{R}^{n \times n}$ such as [8]:

$$P_i = \frac{1}{T_s - 1} X_i X_i^T \quad (1)$$

"*The diagonal elements hold the variance of the signal at each electrode and the off-diagonal elements hold the covariance between all electrode pairs. The SCM does not contain any temporal information at all but contains all spatial information*" [9].

Consider the case $n = 2$. Let $x_1(t)$ and $x_2(t)$ be the two EEG time series recorded at electrodes C3 and C4. For the $i$-th trial, the SCM is [9]:

$$P_i = \begin{pmatrix} Var(x_{1\,i}) & Cov(x_{1\,i},\ x_{2\,i}) \\ Cov(x_{2\,i},\ x_{1\,i}) & Var(x_{1\,i}) \end{pmatrix} \quad (2)$$

Covariance matrices are Symmetric Positive-Definite (SPD) matrices. Since $Cov(x_{1\,i},\ x_{2\,i}) = Cov(x_{2\,i},\ x_{1\,i})$, $2 \times 2$ SPD matrices can be represented as points in $\mathbb{R}^3$, where each point represents a trial of length $T_s$.

The Riemannian distance between two SPD matrices $P_1$ and $P_2$ is given by:

$$\delta_R(P_1, P_2) = \left\| Log(P_1^{-1} P_2) \right\|_F = \left[ \sum_{i=1}^{n} log^2 \lambda_i \right]^{1/2} \quad (3)$$

where $\lambda_i,\ i = 1 \dots n$ are the real eigenvalues of $P_1^{-1} P_2$.

The geometric mean in the Riemannian sense of m given SPD matrices is defined as:

$$\mathfrak{G}(P_1, \dots, P_m) = \underset{P}{\operatorname{argmin}} \sum_{k=1}^{m} \delta_R^{\ 2}(P, P_k) \quad (4)$$

Finally, the shortest way between two SPD matrices in the Riemannian space of SPD matrices is defined by the geodesic $\gamma(t)$ with $t \in [0,1]$ [10]:

$$\gamma(t) = P_1^{1/2} \left( P_1^{-1/2} P_2 P_1^{-1/2} \right)^T P_1^{1/2} \quad (5)$$

The Minimum Distance to Riemannian Mean (MDRM) then simply computes the Riemannian distance of the trial to classify covariance matrix to the geometric mean of each class.

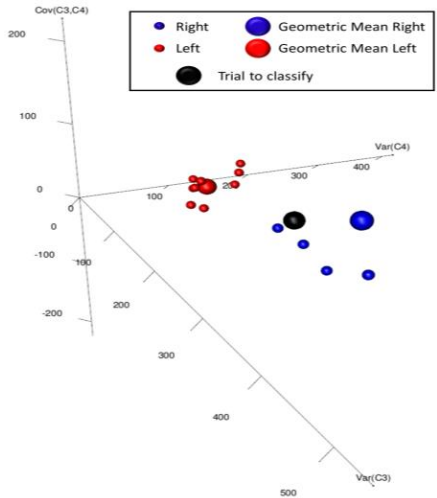

Figure 3. Minimum Distance to Riemannian Mean (MDRM) classification

Fig. 3 shows an example of 2 x 2 covariance matrices for "Left" and "Right" MI classes. The geometric mean for each class is also represented along with a trial to classify. For this example, the trial will be straightforwardly labeled as "Right" (which is the correct label).

## III. PRELIMINARY RESULTS

Firstly, a blinks distribution analysis is carried out to highlight the inter- and intra-subject variability. The Classification Accuracy (CA) obtained from the MDRM method is followingly used as a metric to validate the effectiveness of ABC.

### A. Blinks Distribution

A blinks distribution analysis has been carried out for each subject and each session with the main characteristics summarized in Table I.

TABLE I. BLINKS DISTRIBUTION ANALYSIS

| | Subject A | Subject B | Subject C |
|---|---|---|---|
| Total number of blinks | 112 | 918 | 167 |
| Biggest bin class | [140 160] µV | [180 200] µV | [220 240] µV |
| Percentage of blinks during MI | 25.9 % | 31.8 % | 29.3 % |

The total number of blinks as well as the main bin class vary greatly across subjects. However, the percentage of blinks that arise during MI mental task is noteworthily constant across subjects. This means that without any instruction to refrain from blinking, between a fourth and a third of trials are contaminated with blinks.

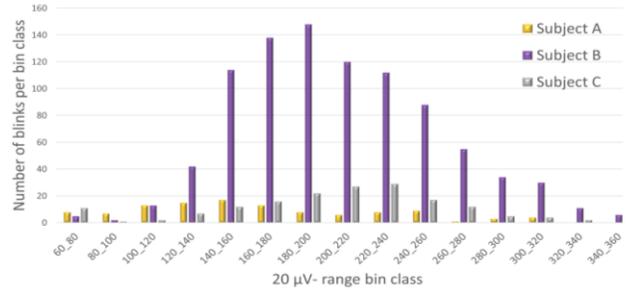

Figure 4. Blinks Distribution across 20 µV wide bin classes

Fig. 4 shows the blinks repartition per bin class for each subject.

### B. Classification and Cross-Validation

In this preliminary work, only 2 EEG channels (C3 and C4) are used for classification with MDRM. The main advantages are (1) the low computational load and (2) the capacity of visualization to assess the validity of this method. Several filters (lowpass bandpass between [8 30] Hz) have been tested with no significant difference on the CA. Consequently, no filters are used before classification with MDRM.

On the other hand, the choice of the Time Interval Restriction (TIR) after the cue instructing the user to perform the mental task has a significant influence over the CA. This time interval seems to be correlated with the "Degree of Alertness" for this task, though the number of subjects is not sufficient yet to infer anything.

TABLE II. TIME INTERVAL RESTRICTION ANALYSIS

| Subject | Session | DOA at start | Task Order | DOA after MI | Length TIR | TIR Start | TIR End |
|---|---|---|---|---|---|---|---|
| Subject A | Sess01 | Slightly tired | 1st | Extreme fatigue | 1.5 | 2 | 3.5 |
| | Sess02 | Slightly tired | 1st | Slightly tired | 1 | 0.5 | 1.5 |
| | Sess03 | Rested | 4th | Rested | 1 | 0 | 1 |
| Subject B | Sess01 | Slightly tired | 2nd | Moderate fatigue | 2 | 0 | 2 |
| | Sess02 | Slightly tired | 2nd | Slightly tired | 1 | 1.5 | 2.5 |
| | Sess03 | Slightly tired | 2nd | Slightly tired | 1.5 | 1 | 2.5 |
| Subject C | Sess01 | Rested | 2nd | Slightly tired | 1 | 0.5 | 1.5 |
| | Sess02 | Slightly tired | 3rd | Moderate fatigue | 2.5 | 1 | 3.5 |
| | Sess03 | Slightly tired | 3rd | Extreme fatigue | 2 | 0 | 2 |

Table II shows the relation between the time interval restriction and the "Degree of Alertness" (DOA) reported by the subject at the end of the MI task.

For now, these time intervals restrictions are chosen manually by comparing the DOA to the CA. In future work, effort will be put into the automatic computation of these time intervals.

### C. Comparison with or without ABC

In this preliminary work, the available dataset is quite small (40 trials per subject and per session). Consequently, a 4-fold cross-validation has been chosen to be applied on all data, since the data cannot be divided into too many folds. Classification Accuracy are thus computed using a simple MDRM as the processing method. Fig. 5 shows the improvement (in percentage) when using ABC as a preprocessing method.

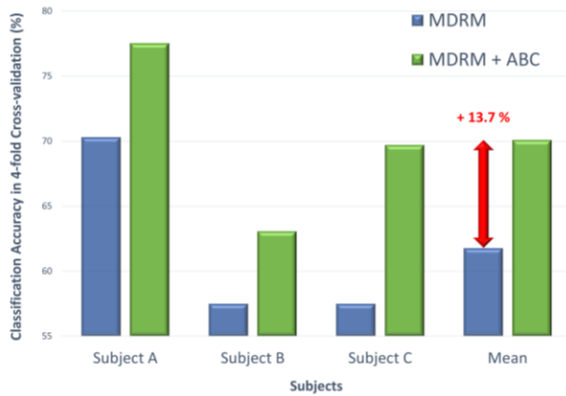

Figure 5. Classification Accuracy Improvement when using ABC with a 4-fold cross-validation

The mean Classification Accuracy shows an improvement of 13.7 % when using ABC as a preprocessing method. Particularly, this improvement ranges from 10 % to 24 % depending on the subject.

## IV. DISCUSSION

This preliminary work shows promising results to use ABC as a preprocessing method. The classification algorithm (MDRM) is here overly simplified (only two EEG channels used for classification) to allow for a clear understanding and visualization of the whole process. Further work will test ABC with a greater number of electrodes for Riemannian geometry classification, as well as Common Spatial Pattern with Linear Discriminant Analysis.

ACKNOWLEDGMENT

Authors would like to thank M. E. M. Mashat for his assistance in conducting the experiment.